\documentclass[preprint,authoryear,12pt]{elsarticle}

\usepackage{epsfig}

\usepackage{amssymb}

\usepackage[ps2pdf,%
a4paper=true,%
breaklinks=true,%
colorlinks=true,%
]{hyperref}

\journal{Advances in Space Research}

\begin{document}

\begin{frontmatter}



\title{The pGAPS experiment: an engineering balloon flight of prototype GAPS}


\author[JAXA]{H.~Fuke\corref{cor}}
\ead{fuke.hideyuki@jaxa.jp}
\author[UCLA]{R.A.~Ong}
\author[Columbia]{T.~Aramaki}
\author[JAXA]{N.~Bando}
\author[UCB]{S.E.~Boggs}
\author[UCB]{P.v.~Doetinchem}
\author[Columbia]{F.H.~Gahbauer}
\author[Columbia]{C.J.~Hailey}
\author[Columbia]{J.E.~Koglin}
\author[Columbia]{N.~Madden}
\author[UCLA]{S.A.I.~Mognet}
\author[Columbia]{K.~Mori}
\author[JAXA]{S.~Okazaki}
\author[Columbia]{K.M.~Perez}
\author[JAXA]{T.~Yoshida}
\author[UCLA]{J.~Zweerink}

\address[JAXA]{Institute of Space and Astronautical Science, Japan Aerospace Exploration Agency (ISAS/JAXA), Sagamihara, Kanagawa 252-5210, Japan}
\address[UCLA]{Department of Physics and Astronomy, University of California, Los Angeles, CA 90095, USA}
\address[Columbia]{Columbia Astrophysics Laboratory, Columbia University, New York, NY 10027, USA}
\address[UCB]{Space Sciences Laboratory, University of California, Berkeley, CA 94720, USA}

\cortext[cor]{Corresponding author}


\begin{abstract}
The General Anti-Particle Spectrometer (GAPS) project is being carried out to search for primary cosmic-ray antiparticles especially for antideuterons produced by cold dark matter. 
GAPS plans to realize the science observation by Antarctic long duration balloon flights in the late 2010s. 
In preparation for the Antarctic science flights, an engineering balloon flight using a prototype of the GAPS instrument, ``pGAPS", was successfully carried out in June 2012 in Japan to verify the basic performance of each GAPS subsystem. 
The outline of the pGAPS flight campaign is briefly reported. 
\end{abstract}


\end{frontmatter}

\parindent=0.5 cm

\section{Introduction}\label{sec:introduction}
Understanding the nature and origin of dark matter is one of the most important goals of 21st century physics\citep{Bertone2004}. 
While we know that the matter makeup of the universe is dominated by dark matter, its nature still remains unsolved. 
The leading candidate for dark matter, or cold dark matter specifically, is the weakly interacting massive particle (or WIMP). 

Three main experimental approaches have been promoted to reveal the nature of dark matter. 
The first is the direct search for dark matter by trying to detect the recoil nucleus from its elastic scattering with target material. 
The second approach is trying to produce dark matter candidate particles by high energy collisions at particle accelerators such as the Large Hadron Collider (LHC). 
And the third approach is the search for indirect evidence of dark matter in the cosmic radiation, such as characteristic signals in gamma-rays, positrons, neutrinos, antiprotons and antideuterons, which could be produced from pair annihilation of WIMPs. 

Of these indirect probes, undiscovered antideuterons are expected to provide a particularly sensitive indirect signature of dark matter, as first pointed out by \citet{Donato2000}. 
The flux of secondary antideuterons produced by the cosmic-ray interactions in the interstellar medium must be kinematically suppressed in the low-kinetic-energy region below around 1~GeV. 
On the other hand, the energy spectrum of primary antideuterons originating from WIMPs can have a softer spectrum through hadronization and coalescence, resulting in a peak above the secondary antideuteron spectrum in the low energy region. 
Therefore, a search for low-energy antideuterons can have essentially a very high signal to background ratio. 
It is pointed out that this attractive ``background-free" feature is a unique advantage of cosmic-ray antideuterons and is common in various dark matter models that produce candidate particles such as the neutralino (i.e. the lightest supersymmetric particle, or LSP), the Kaluza-Klein particle (LKP) and the right-handed neutrino (LZP) in universal extra dimension (UED) theories \citep{Edsjo2004,Profumo2004,Baer2005,Ibarra2009, Brauninger2009}. 


In addition, the antideuteron search is complementary to direct and other indirect detection methods because it probes different portions of the allowed parameter space for dark matter models \citep{Baer2005}. 
In this way, multiple detection methods should be pursued to limit (or even confirm) the available parameter space of the various models. 
Alternatively, the antideuteron search can also probe the signal of primoridial black hole evaporations \citep{Barrau2003}. 

\begin{figure}[htpb]
  \begin{center}
    \includegraphics*[width=14cm]{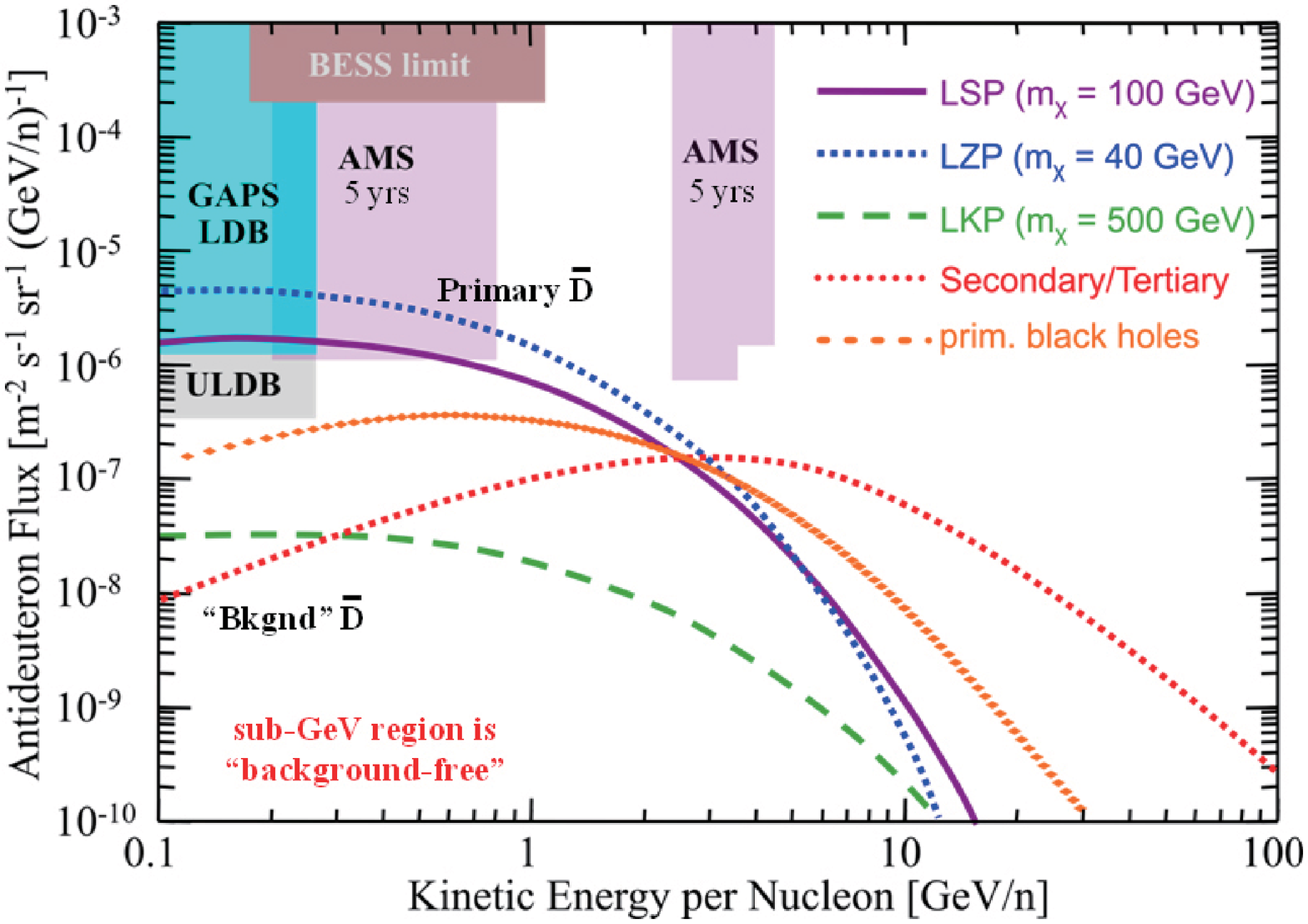}
  \end{center}
  \caption{Cosmic-ray antideuteron energy spectra at the top of the atmosphere expected from various origins, compared with the GAPS sensitivity, the AMS-02 sensitivity, and the BESS upper limit \citep{Fuke2005}. The blue dotted line (LZP), purple solid line (LSP), and green dashed line (LKP) represent the primary antideuteron fluxes originated from the dark matter annihilations \citep{Baer2005}. The orange dotted line represents the primary antideuteron flux originated from primordial black hole evaporations \citep{Barrau2003}. The red dotted line represents the ``background" of secondary/tertiary flux due to the cosmic-ray interactions \citep{Duperray2005}. The solar modulation parameter \citep{Fisk1971} is assumed as $\phi=800~\mbox{MV}$ in these theoretical spectra. The GAPS sensitivity is estimated on the assumption of integrated flight duration of 60 days for LDB and 300 days for ULDB. The AMS-02 sensitivity assumes 5 years observation. }
  \label{fig:dbarspectrum}
\end{figure}

The General Anti-Particle Spectrometer (GAPS) project is a novel approach for the search for low-energy cosmic-ray antideuterons. 
By long duration balloon (LDB) or ultra-long duration balloon (ULDB) flights over Antarctica in the late 2010s, GAPS will probe antideuterons at very high sensitivity, which is more than two or three orders of magnitude better than the only upper limit set by BESS \citep{Fuke2005} or comparable to or even more sensitive than the ongoing AMS-02 experiment on the International Space Station \citep{AMS02}. 
In Fig.~\ref{fig:dbarspectrum}, the GAPS sensitivity for antideuterons is compared with energy spectra expected from several kinds of dark matter.

\section{GAPS project}
\begin{figure}[htpb]
  \begin{center}
    \includegraphics*[width=10cm]{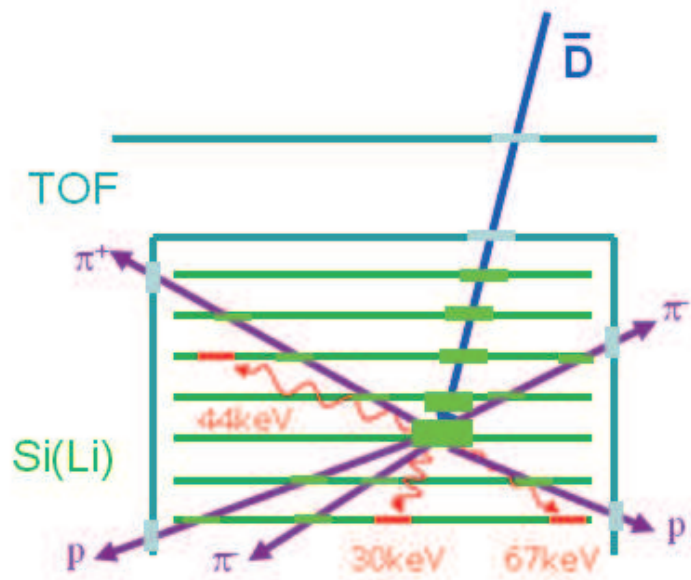}
  \end{center}
  \caption{The detection concept of GAPS. The antiparticle slows down and stops in the target, then forms the excited exotic atom, followed by deexcitation with X-ray emission and nuclear annihilation with pions and protons production. }
  \label{fig:dbaridentification}
\end{figure}

Unlike BESS and AMS-02, which use magnetic mass spectrometers, GAPS introduces an original method that utilizes the deexcitation sequence of exotic atoms \citep{Mori2002,Hailey2004}. 
As an antideuteron arrives from space, it is slowed down by the $\mbox{d}E/\mbox{d}x$ energy losses in the residual atmosphere, in the GAPS time-of-flight counters, and in the target material. 
Just after stopping in the target, the antideuteron forms an exotic atom in an excited state with near unity probability. 
Then, through radiative transitions, the exotic atom deexcites with the emission of characteristic X-rays. 
The energy of the X-ray is strictly determined by the physics of the exotic atom, and thus provides particle identification information of the incoming (anti)particle species. 
After the X-ray emission, the antideuteron annihilates in the nucleus, producing a characteristic number of pions and protons, which provides additional particle identification information. 
By requiring the coincidence of X-rays and pions or protons, we can distinguish antideuterons from background processes including cosmic-ray antiprotons, protons, and accidental X-rays. 
The usefulness of this detection principle (Fig.~\ref{fig:dbaridentification}) was verified and confirmed through accelerator tests with various target materials using the KEK antiproton beam-line \citep{Hailey2006}. 


The GAPS detector being developed mainly consists of a large pixelated lithium drifted silicon detectors, Si(Li), surrounded  by large Time-of-Flight (TOF) counters \citep{Fuke2008,Aramaki2010}. 
In a cuboidal space about 2~m on a side, thousands of palm-size Si(Li) wafers are arranged in a dozen layers. 
The Si(Li) detector serves as a degrader, a depth sensing detector to know the penetration length, a stopping target, an X-ray energy measurement detector and a charged particle tracker. 
The required energy resolution for Si(Li) is around 3~keV, which is well achievable at low temperatures around -30$^\circ$C. 

The TOF counters with 3 -- 5~mm thick plastic scintillator paddles generate the trigger signal and measure the time-of-flight (or velocity) and the energy deposit $\mbox{d}E/\mbox{d}x$. 
The TOF counters can roughly determine the arrival direction, and work as a pion/proton detector and a veto counter as well. 
The required time resolution for the TOF is around 0.5~nsec. 

As previously described, GAPS plans to realize the antideuteron search by LDB or ULDB balloon flights over Antarctica. 
A polar balloon flight is optimal for GAPS, because the low rigidity cutoff near the geomagnetic pole allows low-energy charged cosmic rays to be observed directly. 

\section{Engineering flight ``pGAPS"}
Preparatory for the GAPS science flight, we carried out ``pGAPS", a balloon experiment with a prototype of GAPS instruments \citep{Hailey2011}. 
pGAPS aims to verify and demonstrate the basic performance of each GAPS subsystem in balloon-flight conditions. 
The pGAPS flight has been prepared to be launched from Taiki Aerospace Research Field (TARF), a JAXA's scientific balloon facility at Taiki, Hokkaido \citep{Fuke2010}. 
pGAPS cannot detect antiprotons nor antideuterons for the following three reasons: (i) high geomagnetic rigidity cutoff (around 8~GV for 50\% cutoff), (ii) short flight time (around 3 hours at level altitude), and (iii) a small detector acceptance (around 0.054~m$^2$sr). 

Instead, there were three goals for pGAPS: (i) to demonstrate the operation of the Si(Li) detectors and the TOF counters at float altitude and ambient temperature, (ii) to obtain thermal data to verify the Si(Li) cooling system, and (iii) to measure the incoherent background level. 

Especially for the second goal, the actual balloon flight is essential, because in the lab it is not easy to simulate all at once the flight conditions with low temperatures, low pressure (but not vacuum), solar radiation, and infrared radiation from the earth. 
The first and the third goals are important to confirm the noise level measured at the preflight environmental testing as well as to verify the background level estimated in the detector design simulations.

\subsection{pGAPS payload}
\begin{figure}[htpb]
  \begin{center}
    \includegraphics*[width=14cm]{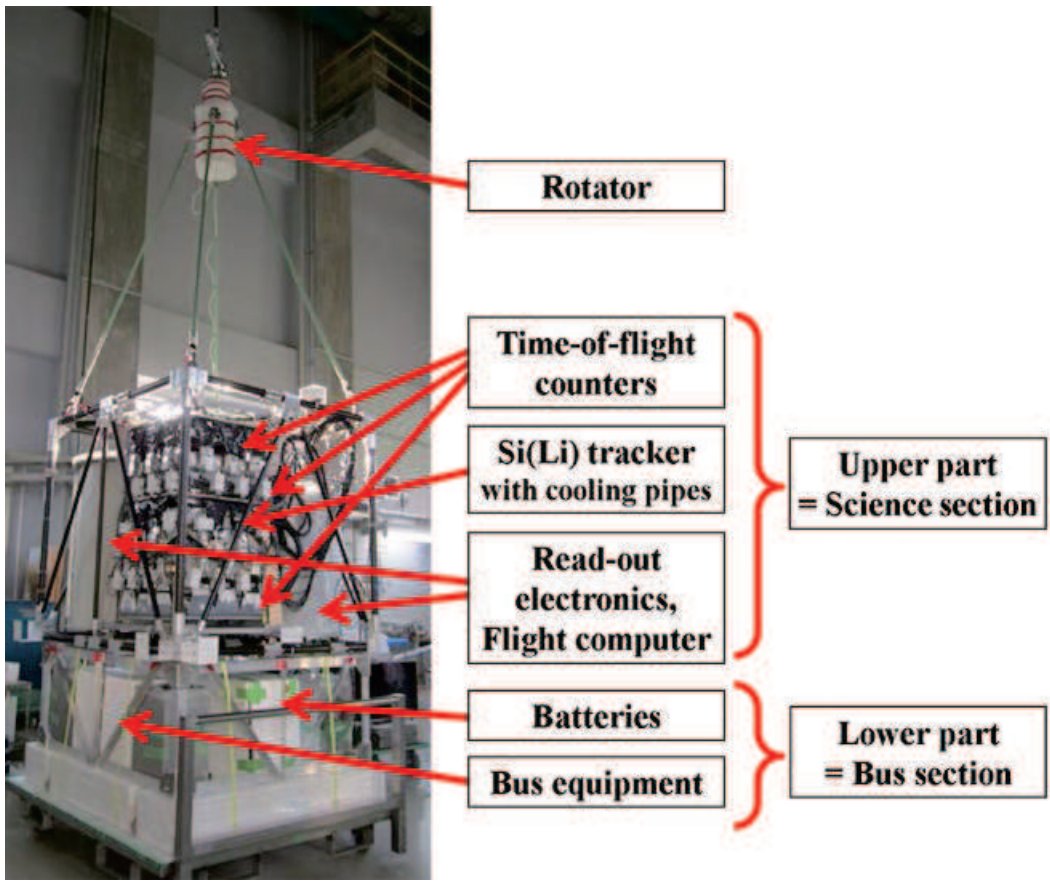}
  \end{center}
  \caption{Configuration of the pGAPS payload.}
  \label{fig:payloadconfiguration}
\end{figure}

Figure~\ref{fig:payloadconfiguration} shows the overall configuration of the pGAPS payload. 
The pGAPS payload consisted of two sections: the upper section and the lower section. 
In the lower section, bus equipment for JAXA's balloon operation as well as batteries were mounted. 
Most of the pGAPS science instruments, such as Si(Li) detectors, TOF counters, and front-end electronics, were mounted in the upper section. 
Above the top of the payload (beneath the parachute flight train), a rotator was attached in order to roughly control the payload attitude. 
The pGAPS payload weighed about 510~kg, and the total power consumption of the pGAPS science instruments was about 430~W. 

On pGAPS, six Si(Li) detectors (wafers) were mounted, arranged in two columns of three detectors each. 
Each Si(Li) detector had a preamplifier with individual readout for each eight strips with dual energy ranges for X-rays and pions/protons. 
The energy resolution for X-rays at -35$^\circ$C measured prior to the pGAPS flight with an Americium 241 radioactive source put on top of the fully assembled payload was about 5.6~keV. 
For the in-flight calibration of the Si(Li) detectors, an X-ray tube (with filters to shape peaks at 26~keV and 35~keV) was also mounted on pGAPS.  

For the cooling system of the Si(Li) detectors, two kinds of prototype cooling systems were mounted. 
One was the baseline plan using a closed-loop forced-convection system with a pumped fluid, which was actually utilized for cooling the Si(Li) detectors during the flight. 
The other was a challenging optional plan using an oscillating heat pipe (OHP) system, which was attached to a dummy heat load independent from the actual detectors. 
Both systems transferred the heat to a separate radiator attached on the payload sidewall to dissipate the heat towards space. 

The TOF counter mounted on pGAPS consisted of three layers of crossed plastic scintillator paddles. 
Each scintillator paddle had a photomultiplier tube (PMT) readout from both ends. 
Both the PMT anode and dynode signals were buffered by a high-speed amplifier and sent to the readout electronics. 
Leading-edge discriminators sampled the inverted dynode signal and stopped time of digital converters (TDCs) with 50~ps resolution. 
The achieved time resolution was around 0.5~ns. 
The anode signal was sampled by a charge-ADC (qADC) with 0.25~pC resolution. 
In this way, the TOF signal was used for the $\mbox{d}E/\mbox{d}x$ measurement, the time-of-flight measurement, and triggering. 

As for the trigger, two types of trigger modes were implemented: 
TOF trigger mode and Si(Li) self-trigger mode. 
In order to adapt to these two trigger modes, the Si(Li) readout system was designed to operate in two corresponding modes. 
The main mode was to start the event processing on the TOF trigger formed by the TOF logic to measure the charged particle track in coincidence with all sub-detectors. 
The Si(Li) self-trigger mode was used for X-ray calibrations and background measurements and ignored the TOF trigger when the Si(Li) self-trigger was issued. 

Data from all subsystems were collected by an onboard flight computer. 
The data were stored in onboard data storage and also were sent to the ground by telemetry via the bus equipment. 
Tele-commands could be sent from the ground to the flight computer to enable payload controls including software mode switching and power on/off switching of each subsystem. 
The electronics, such as the flight computer and the front-end electronics both for the Si(Li) detectors and the TOF counters were installed in two vessels --- one pressurized and one watertight --- to survive the different conditions encountered during the mission: low atmospheric pressure at high altitude and sea water at recovery. 

As the power source, primary lithium batteries were used for pGAPS to reduce the payload weight. 
GAPS for Antarctic flight will use solar panels and rechargeable batteries. 
The power from the lithium battery were distributed to each subsystem by DC-DC converters. 
The details of the pGAPS payload are to be reported elsewhere.

\subsection{pGAPS campaign}
The upper section and the lower section of the pGAPS payload were integrated at UC Berkeley and at ISAS/JAXA, respectively. 
In May 2012, the upper section was transported to Japan for the full payload integration and ground tests of the overall pGAPS system. 
The preparation in Japan was carried out at first at the ISAS/JAXA Sagamihara campus. 
Then, after moving to the launch site TARF, the final preparations, including telecommunication compatibility tests and a dress rehearsal, were carried out. 

The pGAPS payload was launched at 4:55 AM on June 3rd from TARF. 
After a 3-hour ascent, of which 1.5 hours were spent drifting eastward according to the so-called boomerang flight operation \citep{Nishimura1981}, the balloon reached the level altitude at around 8:05 AM. 
Then, the balloon floated westward for 3 hours at the level altitudes of around 31 - 33~km.
The flight was terminated near the coast, and then the payload was safely recovered from the sea within only 9 minutes after the splash down at 11:36 AM. 

\begin{figure}[htpb]
  \begin{center}
    \includegraphics*[width=12cm]{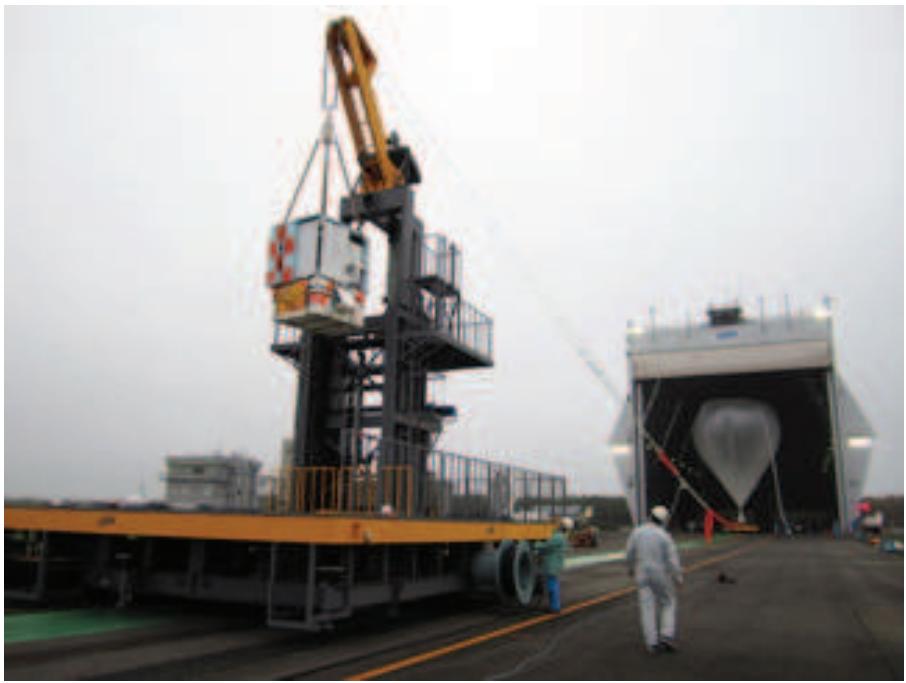}
  \end{center}
  \caption{The pGAPS payload set on the launcher during the balloon inflation.}
  \label{fig:launch}
\end{figure}

\begin{figure}[htpb]
  \begin{center}
    \includegraphics*[width=12cm]{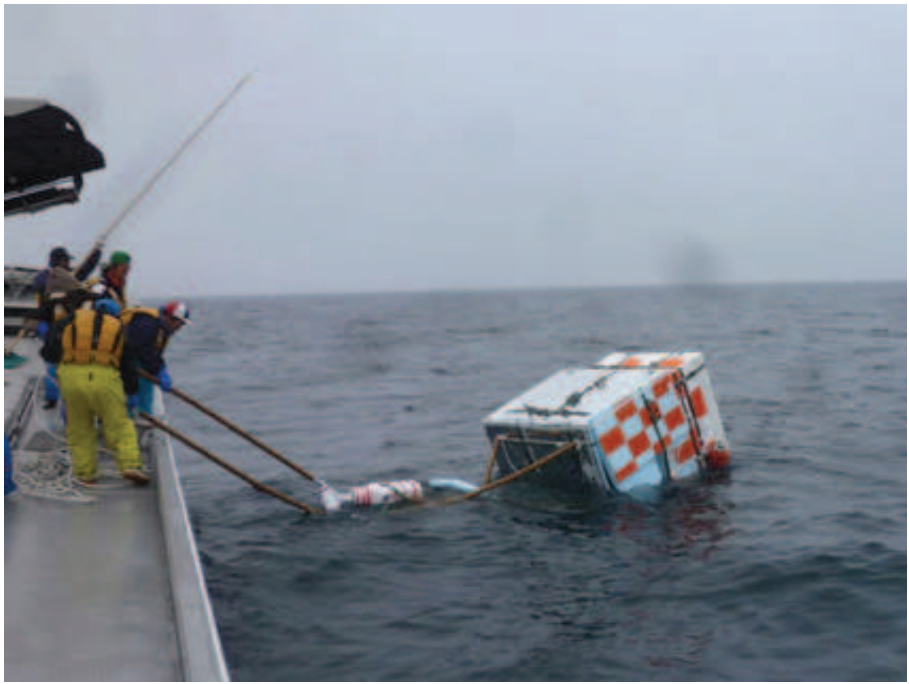}
  \end{center}
  \caption{The pGAPS payload recovered on the sea.}
  \label{fig:recovery}
\end{figure}

\begin{figure}[htpb]
  \begin{center}
    \includegraphics*[width=12cm]{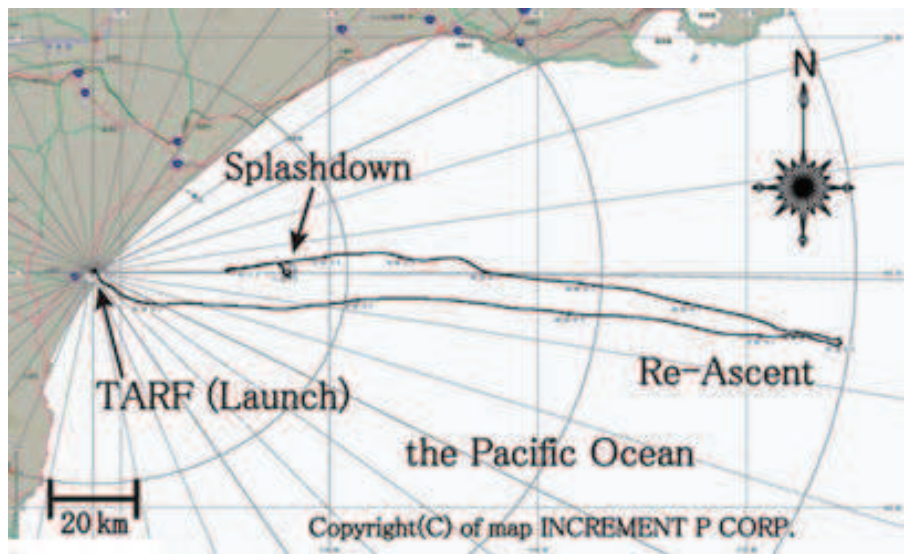}
  \end{center}
  \caption{Trajectory of the pGAPS balloon flight. }
  \label{fig:trajectory}
\end{figure}

\begin{figure}[htpb]
  \begin{center}
    \includegraphics*[width=14cm]{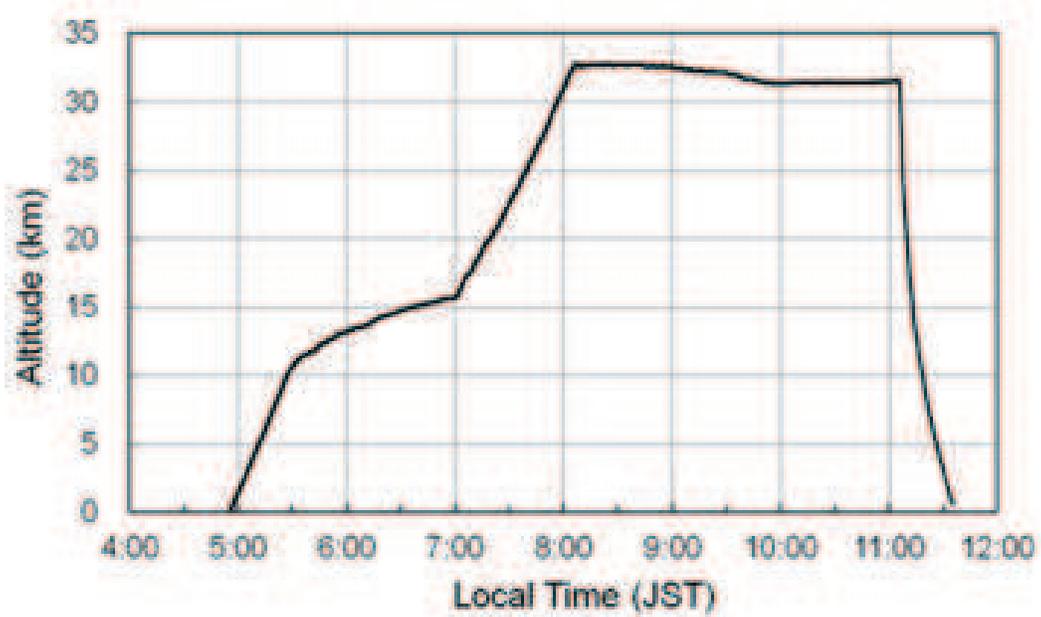}
  \end{center}
  \caption{Altitude profile of the pGAPS balloon flight.}
  \label{fig:altitude}
\end{figure}

Figures~\ref{fig:launch} and \ref{fig:recovery} show the pGAPS payload just before the launch and during the recovery. 
Figures~\ref{fig:trajectory} and \ref{fig:altitude} show the flight trajectory and the flight altitude profile, respectively.

\subsection{pGAPS flight data}
\begin{figure}[htpb]
  \begin{center}
    \includegraphics*[width=12cm]{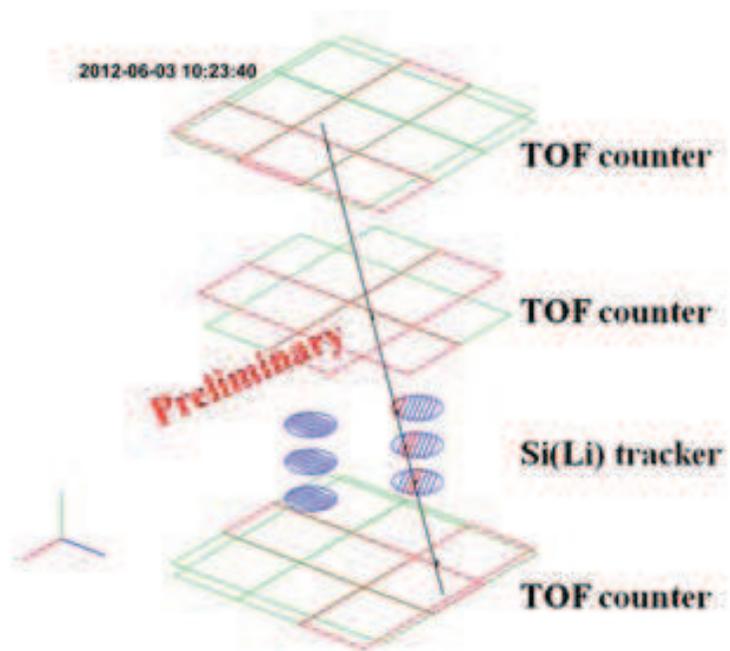}
  \end{center}
  \caption{An example of an event recorded during the pGAPS flight.}
  \label{fig:eventdisplay}
\end{figure}

The pGAPS payload was operated throughout the flight. 
More than one million events were recorded. 
Unfortunately the rotator for the payload attitude control failed due to an operation mistake during the flight. 
However all the other payload components basically worked well throughout the flight. 
Our preliminary analysis has confirmed that the Si(Li) detectors and the TOF counters showed good, stable performance with low noise. 
Figure~\ref{fig:eventdisplay} shows an example of an event recorded during the flight.  
Temperature data to evaluate the cooling system and incoherent cosmic-ray background data in flight-like configuration were also obtained as planned. 
The result of the ongoing analyses will be presented elsewhere. 

\section{Summary}
The GAPS project aims to search for undiscovered cosmic-ray antideuterons produced by cold dark matter annihilations. 
GAPS plans to fly over Antarctica multiple times by LDB (or ULDB) balloon flights to achieve a high antideuteron sensitivity. 
In preparation for the Antarctic flights, a prototype of GAPS, ``pGAPS", was launched from a Japanese balloon base in June 2012 to verify the basic performance of the GAPS subsystems. 
The pGAPS flight was very successful; all the key components worked well and more than one million events were recorded during the flight. 
Detailed analyses of the flight data is now ongoing.

\section*{Acknowledgments}
We thank J.~Hoberman for the development of the GAPS electronics, and we also thank G.~Tajiri and D.~Stefanik for the mechanical engineering support. 
We are grateful for the the support of Prof.~H.~Ogawa and Prof.~Y.~Miyazaki for the OHP development. 
We thank the Scientific Balloon Office of ISAS/JAXA for the professional support of the pGAPS flight. 
This work is partly supported in the US by NASA APRA Grants (NNX09AC13G, NNX09AC16G) and in Japan by MEXT grants KAKENHI (22340073).



\newpage

\newpage

\newpage

\newpage

\newpage

\newpage

\newpage

\newpage

\begin{thebibliography}{}


\bibitem[AMS-02(website)]{AMS02}
AMS-02 website, http://www.ams02.org/ .

\bibitem[Aramaki et al.(2010)]{Aramaki2010}
Aramaki, T., Boggs, S.E., Craig, W.W, {\em et al.}, 
Antideuterons as an indirect dark matter signature: Si(Li) detector development and a GAPS balloon mission, 
Adv. Space Res. 46, 1349-1353, 2010. 

\bibitem[Baer and Profumo(2005)]{Baer2005}
Baer, H., Profumo, S, 
Low energy antideuterons; shedding light on dark matter, 
J. Cosmol. Astropartic. Phys. 12, 008, 2005. 

\bibitem[Barrau et al.(2003)]{Barrau2003}
Barrau, A., Boudoul, G., Donato, F., {\em et al.}, 
Antideuterons as a probe of primordial black holes, 
Astron. Astrophys. 398, 403, 2003. 

\bibitem[Bertone(2004)]{Bertone2004}
Bertone, G., Hooper, D., and Silk, J., 
Particle Dark Matter: Evidence, Candidates and Constraints, 
Phys. Rep. 405, 279-390, 2005. 

\bibitem[Br$\ddot{\mbox{a}}$uninger and Cirelli(2009)]{Brauninger2009}
Br$\ddot{\mbox{a}}$uninger, C.B.,  Cirelli, M. 
Antideuterons from heavy dark matter, 
Phys. Lett. B 678, 20-31, 2009. 

\bibitem[Donato et al.(2000)]{Donato2000}
Donato, F., Fornengo, N., Salati, P., 
Antideuterons as a signature of supersymmetric dark matter, 
Phys. Rev. D 62, 043003, 2000. 

\bibitem[Duperray et al.(2005)]{Duperray2005}
Duperray, R., Baret, B., Maurin, D., {\em et al.}, 
Flux of light antimatter nuclei near Earth, induced by cosmic rays in the Galaxy and in the atmosphere, 
Phys. Rev. D 71, 083013, 2005. 

\bibitem[Edsjo et al.(2004)]{Edsjo2004}
Edsjo, J., Schelke, M., Ullio, P., 
Direct versus indirect detection in mSUGRA with self-consistent halo models, 
J. Cosmol. Astropartic. Phys. 09, 004, 2004. 

\bibitem[Fisk(1971)]{Fisk1971}
Fisk, L. A., 
Solar modulation of galactic cosmic rays, 
J. Geophys. Res. 76, 221-226, 1971. 

\bibitem[Fuke et al.(2005)]{Fuke2005}
Fuke, H., Maeno, T., Abe, K., {\em et al.}, 
Search for cosmic-ray antideuterons, 
Phys. Rev. Lett. 95, 081101, 2005. 

\bibitem[Fuke et al.(2008)]{Fuke2008}
Fuke, H., Koglin, J.E., Yoshida, T., {\em et al.}, 
Current status and future plans for the general antiparticle spectrometer (GAPS), 
Adv. Space Res. 41, 2056-2060, 2008. 

\bibitem[Fuke et al.(2010)]{Fuke2010}
Fuke, H., Akita, D., Iijima, I., {\em et al.}, 
A new balloon base in Japan, 
Adv. Space Res. 45, 490-497, 2010. 

\bibitem[Hailey et al.(2004)]{Hailey2004}
Hailey, C.J., Craig. W.W., Harrison, F.A., {\em et al.}, 
Development of the gaseous antiparticle spectrometer for space-based antimatter detection, 
Ncul. Instr. Meth. B 214, 122, 2004. 

\bibitem[Hailey et al.(2006)]{Hailey2006}
Hailey, C.J., Aramaki, T., Craig. W.W., {\em et al.}, 
Accelerator testing of the general antiparticle spectrometer, a novel approach to indirect dark matter detection, 
J. Cosmol. Astropartic. Phys. 0601, 007, 2006. 

\bibitem[Hailey et al.(2011)]{Hailey2011}
Hailey, C.J., Aramaki, T., Boggs, S.E., {\em et al.}, 
Antideuteron based dark matter search with GAPS: Current progress and future prospects, 
Adv. Space Res. (2011), doi:10.1016/j.asr.20011.04.025 (in press). 

\bibitem[Ibarra and Tran(2009)]{Ibarra2009}
Ibarra, A., Tran, D., 
Antideuterons from dark matter decay, 
J. Cosmol. Astropartic. Phys. 0906, 004, 2009. 

\bibitem[Mori et al.(2002)]{Mori2002}
Mori, K., Hailey, C.J., Baltz, E.A., {\em et al.}, 
A novel antimatter detector based on X-ray deexcitation of exotic atoms, 
Astrophys. J. 566, 604, 2002. 

\bibitem[Nishimura and Hirosawa(1981)]{Nishimura1981}
Nishimura, J., Hirosawa, H., 
Systems for long duration flights, 
Adv. Space Res. 1, 239-249, 1981. 

\bibitem[Profumo and Ullio(2004)]{Profumo2004}
Profumo, S., Ullio, P., 
The role of antimatter searches in the hunt for supersymmetric dark matter, 
J. Cosmol. Astropartic. Phys. 0407, 006, 2004. 

\end{thebibliography}
\end{document}